\documentclass[preprint,12pt]{elsarticle}

\usepackage{amssymb}
\usepackage{amsmath}
\usepackage{amsthm}
\newtheorem{proposition}{Proposition}

\journal{Nuclear Physics B}

\begin{document}

\begin{frontmatter}



\title{Temporal Tensors and Quantum Shortcut Dynamics in a Supermaze of Multidimensional Time} 


\author{Koffka Khan} 

\affiliation{organization={The University of the West Indies, Department of Computing and Information Technology},
            city={St. Augustine},
            country={Trinidad and Tobago}}

\begin{abstract}
We develop a theoretical framework that unifies concepts of multiple time dimensions, quantum shortcut dynamics, and complex topological structures (``supermazes'') to explore novel phenomena in quantum and classical systems. In particular, we introduce a \textit{Temporal Tensor Formalism} to describe multidimensional time, define \textit{Quantum Shortcut Operators} that enact near-instantaneous state transitions, and incorporate these into a \textit{supermaze} topological model inspired by labyrinthine geometry and network complexity. We show how this framework can give rise to surprising effects such as anomalous thermodynamic relaxation (analogous to the Mpemba effect) in quantum systems. Theoretical implications for quantum computing (including quantum cloud networks) are discussed, and connections are drawn to established mathematical paradoxes and physical principles. 
\end{abstract}


\begin{highlights}
	\item Proposes a Temporal Tensor Formalism to model multidimensional time evolution in quantum systems.
	\item Introduces Quantum Shortcut Operators that enable near-instantaneous transitions across quantum states.
	\item Develops a topological supermaze framework to analyze complex quantum navigation paths.
	\item Explains anomalous thermodynamic behavior such as the Mpemba effect within a multi-time structure.
	\item Discusses implications for quantum cloud computing and machine learning through supergeometric control strategies.
\end{highlights}

\begin{keyword}
Multidimensional time \sep Quantum shortcut operators \sep Temporal tensor formalism \sep Topological complexity \sep Thermodynamic anomalies


\end{keyword}

\end{frontmatter}



\section{Introduction} The nature of time and its role in physical law remain deeply intriguing. Standard physics assumes a single temporal dimension, yet speculative theories have posited the existence of multiple independent time dimensions [1]. \textit{Two-time physics}, for example, extends spacetime with an extra timelike dimension, providing a higher-dimensional symmetry that unifies various one-time physical formulations [1]. On the other hand, arguments from cosmology suggest that more than one time dimension could lead to fundamental instabilities and unpredictability [2]. This tension between exotic mathematical possibility and physical plausibility forms a key motivation for our work. We ask: \emph{What new structures or laws might emerge if one considers multiple temporal directions, and can these be harnessed without violating consistency?} A parallel inspiration comes from advances in quantum computing and control. It is known that quantum algorithms can sometimes achieve dramatic speedups by effectively taking “shortcuts” through computation—exploring many paths in superposition [3]. Recent developments in quantum control, such as \textit{shortcuts to adiabaticity}, enable systems to reach target states much faster than normally allowed by slow adiabatic evolution [6]. These quantum shortcuts hint at a tantalizing concept: if a system can somehow exploit additional degrees of freedom (beyond the usual one-dimensional time flow) or cleverly bypass intermediate states, it might accomplish tasks at unprecedented speeds. However, such shortcuts are constrained by fundamental limits—no operation can violate the laws of thermodynamics or quantum uncertainty. Indeed, physicists have found that any attempt to evolve a quantum system arbitrarily fast incurs a large energy cost, in accordance with quantum speed limit theorems [7]. This raises a paradoxical question: \emph{Can one circumvent temporal limitations without breaking physical laws, perhaps by using novel pathways that ordinary dynamics do not explore?} We draw further insight from classical geometry and thermodynamics. History provides examples where complex problems hide elegant structure. An illustrative case is Descartes’ circle problem (the “kissing circles” puzzle), which asks for the radius of a fourth circle tangent to three given mutually tangent circles. Descartes in 1643 derived a beautiful quadratic formula relating the curvatures of all four circles [4], later popularized by Soddy [5]. This \textit{Descartes circle theorem} revealed that a seemingly intricate geometric arrangement is governed by a simple invariant relation. By analogy, one may suspect that complicated networks or “mazes” of possible states could obey hidden mathematical rules. Another intriguing phenomenon is the \textit{Mpemba effect}, wherein hot water can freeze faster than cold water under certain conditions[10]. Long considered a paradox, this effect suggests that initial state differences can lead to non-intuitive faster relaxations. If such anomalous behavior is possible in classical systems, could analogous effects occur in quantum or multi-dimensional time systems, where an unconventional route in state-space leads to quicker equilibrium? Contributions of this work: Motivated by these ideas and paradoxes, we develop a theoretical framework that combines multiple time dimensions, shortcut dynamics, and complex topology: 
\begin{itemize} \item We introduce a \textbf{Temporal Tensor Formalism} that generalizes time to an $N$-dimensional tensor quantity. This provides a foundation to incorporate more than one time dimension in physical equations while preserving consistency. \item We define \textbf{Quantum Shortcut Operators}, inspired by shortcuts to adiabaticity, as operators or functionals that achieve direct state transitions faster than the normal dynamical evolution. We formalize conditions under which such shortcuts can occur and the trade-offs (such as energy cost or information loss) required by fundamental principles. \item We propose a \textbf{Supermaze Topological Model} to unify these concepts, treating the complex network of states and paths (including multi-time routes and shortcuts) as a topological ``maze''. We draw on concepts from supergeometry and topological complexity to characterize the structure of this maze and quantify the routing problem within it. \item We demonstrate that in this framework, \textbf{thermodynamic anomalies} can arise in quantum systems. In particular, we discuss how a quantum analog of the Mpemba effect might occur: certain excited states can relax faster than some lower-energy states because the availability of shortcut pathways or multi-time dynamics circumvents bottlenecks of the usual thermalization process. 
\end{itemize} In essence, our work provides a novel theoretical synthesis: Section 2 reviews necessary background in multiple time dimensions, quantum control, geometry, and topology. Section 3 develops the mathematical formulation of our model, including fundamental equations and definitions. Section 4 builds the theoretical framework with four key components: temporal tensors, quantum shortcut operators, supermaze topology, and thermodynamic anomalies. In Section 5, we discuss the implications of our results, with an emphasis on potential impact for quantum cloud computing and how our model relates to known physical principles and thought experiments. Finally, Section 6 concludes the paper and outlines directions for future research, such as more rigorous simulations and experimental considerations. 
\section{Preliminaries and Theoretical Background} 
\subsection*{Multidimensional Time and Supergeometry} The possibility of multiple time dimensions has been explored in theoretical physics, albeit controversially. In a spacetime with $N$ temporal dimensions $t_1,\dots,t_N$, and $d$ spatial dimensions, the invariant interval might be generalized as \begin{equation}\label{eq:metric} ds^2 = -\sum_{i=1}^{N} (c_i,dt_i)^2 + \sum_{j=1}^{d} dx_j^2~, \end{equation} where $c_i$ are characteristic speeds (e.g. $c_1=c$ the speed of light) for each time dimension. Models such as two-time physics (the $N=2$ case) embed our $3+1$ dimensional world into a higher-dimensional space with a $(-,-,+,+\dots)$ metric signature [1]. One benefit of such models is an enlarged symmetry algebra that can unify disparate one-time physical laws [1]. For instance, Bars (2001) showed that a suitably constructed two-time field theory in $d+2$ dimensions can yield the usual $d+1$ dimensional dynamics as shadows of a more symmetric underlying structure [1]. However, introducing extra time-like directions is not straightforward. A basic issue is the well-posedness of physical laws: with multiple times, partial differential equations can become \textit{ultrahyperbolic}, lacking a clear initial value formulation. Indeed, it has been proven that without additional constraints, equations like a wave equation with two time variables do not yield deterministic evolution [3]. Craig and Weinstein (2009) demonstrated that to obtain a unique solution, initial data must be prescribed on a special hypersurface and satisfy a hidden consistency condition [3]. In essence, not every specification of fields at an ``initial moment'' of multi-time will lead to a coherent history; the data must lie in a certain subspace (often defined by an integral constraint) to avoid unphysical solutions. This indicates that if multiple time dimensions exist, they might be coupled or “entangled” in a way that preserves causality and determinism. To incorporate multiple times in a controlled manner, we introduce the notion of a \emph{temporal tensor}. In this context, a temporal tensor $T^{a_1 \dots a_N}$ is a mathematical object that carries indices in multiple time directions. The simplest case is a \textit{time vector} $T^a = (t_1, t_2, \dots, t_N)$ representing an event’s coordinates in an $N$-time system. More generally, we can define higher-rank temporal tensors to encode relationships between different time directions. For example, a rank-2 temporal metric $G^{ij}$ could be used to relate intervals along $t_i$ and $t_j$. Eq.~\eqref{eq:metric} effectively contains such a metric (diagonal in the $t_i$ basis). Throughout this paper, we use a superindex notation $t^I$ (with $I=1,\dots,N$) when referring to an $N$-dimensional time coordinate collectively. An extension of these ideas brings us to \textit{supergeometry}. Supergeometry generalizes classical geometry by including anticommuting (Grassmann) coordinates alongside ordinary coordinates. In a \textit{supermanifold}, one has bosonic coordinates (here, spatial $x_j$ and temporal $t_i$) and fermionic coordinates (often denoted $\theta$). The latter can be thought of as mathematical devices that capture discrete or binary degrees of freedom. In our framework, one motivation to consider a superspace $(t^I, x^j | \theta^\alpha)$ is to encode additional structural or combinatorial information about state-space paths. For instance, the presence or absence of a quantum shortcut might be represented by a Grassmann variable (since a fermionic variable squared is zero, it can act like a flag). While a full development of a supersymmetric extension is beyond our scope, we note that formulating our model in a superspace could naturally incorporate features like whether a path is normal'' or a shortcut'' (much as supersymmetry uses Grassmann variables to distinguish particle states). This viewpoint aligns with the concept of a \emph{supermaze} introduced later, hinting that the labyrinth of possible routes might be treated as a supergeometric object. 
\subsection*{Quantum Shortcuts and Speed Limits} A second ingredient of our theoretical background is the notion of quantum shortcuts. In quantum mechanics, the evolution of a state $|\Psi(t)\rangle$ is unitary and governed by the Hamiltonian $H(t)$. The adiabatic theorem tells us that if $H(t)$ changes slowly, a system initially in an eigenstate of $H$ will remain in the instantaneous eigenstate. \textit{Shortcuts to adiabaticity} (STA) are protocols that modify the evolution so the final outcome of a slow, adiabatic process can be achieved in a much shorter time [6]. These protocols often involve adding a carefully chosen auxiliary Hamiltonian or driving term $H_{\text{CD}}(t)$ (counter-diabatic term) to guide the state along the adiabatic path without needing slow evolution. STA methods have been demonstrated in various quantum systems, effectively allowing nearly instantaneous population transfer, quantum state preparation, or cooling, all while avoiding excitations that a naive fast ramp would induce [6]. With the promise of such quantum shortcuts comes the question of fundamental limits. Can one make a quantum transition arbitrarily fast? The answer is constrained by the \emph{quantum speed limit}, a set of bounds arising from the time-energy uncertainty principle. In simple terms, there is a minimum time $\Delta \tau$ required for a quantum system to evolve between two distinguishable states given a certain energy budget. One formulation (Mandelstam–Tamm) is $\Delta \tau \cdot \Delta E \ge \frac{\pi \hbar}{2}$, where $\Delta E$ is the energy uncertainty of the state [7]. Another (Margolus–Levitin) says $\Delta \tau \ge \frac{\pi \hbar}{2E_{\text{avg}}}$ for average energy $E_{\text{avg}}$ above the ground state. These inequalities imply that “fast-forwarding” a quantum evolution is not free – it demands greater energy or broader bandwidth. Campbell and Deffner (2017) explicitly analyzed quantum shortcuts and showed that attempting to speed up a quantum process inevitably incurs an energetic cost, preventing any violation of the second law of thermodynamics [7]. We will make use of these ideas by defining idealized \textit{quantum shortcut operators} in our model and then investigating their limitations. Formally, consider a unitary operator $\hat{U}(t_f,t_i)$ that evolves the system from time $t_i$ to $t_f$ under some Hamiltonian $H$. A “shortcut” operator $\hat{S}$ would be one that yields the same final state $\hat{U}(t_f,t_i)|\Psi(t_i)\rangle$ in a shorter duration $\Delta \tau' < (t_f - t_i)$. In standard quantum theory, $\hat{S}$ can be realized by changing $H(t)$ to $H'(t)$ (for $t_i \le t \le t_i+\Delta\tau'$) such that $\mathcal{T}\exp[-\frac{i}{\hbar}\int_{t_i}^{t_i+\Delta\tau'}H'(t)dt] = \mathcal{T}\exp[-\frac{i}{\hbar}\int_{t_i}^{t_f}H(t)dt]$, where $\mathcal{T}$ denotes time-ordering. The counter-diabatic $H_{\text{CD}}$ mentioned is one choice for $H'(t)$. In our abstraction, we treat $\hat{S}$ as an operator that connects two points in the state space “directly”. A crucial point is that $\hat{S}$ might not be implementable within the original system’s confines—it could require a fundamentally different use of resources (for example, access to an extra dimension or an external control field). By positing multiple time dimensions, one might imagine a scenario where evolution in an auxiliary time $t_2$ facilitates an effective jump in the physical time $t_1$. This is a conceptual leap: using an extra temporal degree of freedom to sidestep the usual sequential progression. While speculative, this resonates with how a traveler in a maze might momentarily step into a higher dimension to bypass walls in the maze, then re-enter the original plane at a new location. We emphasize that any shortcut, either in a single time or multi-time context, must respect Proposition \ref{prop:speedlimit} below. \begin{proposition}[Quantum Speed Limit]\label{prop:speedlimit} No quantum shortcut can circumvent the fundamental time–energy bound. If a shortcut operator $\hat{S}$ achieves a state transformation in effective time $\Delta\tau'$, then one must have $E_{\mathrm{cost}};\Delta\tau' \ge \frac{\pi \hbar}{2}$, where $E_{\mathrm{cost}}$ is the additional energy expended by the shortcut protocol. \end{proposition} \noindent \textit{Proof Sketch.} This statement encapsulates results from quantum speed limit derivations (Mandelstam–Tamm and Margolus–Levitin inequalities) [7]. In essence, $\hat{S}$ can be viewed as driving the system with some perturbed Hamiltonian $H'(t)$ that includes high-frequency components or large instantaneous eigenvalue gaps. The energy cost $E_{\mathrm{cost}}$ can be associated with the norm $|H' - H|$ or the required instantaneous excitation above the original spectrum. The standard quantum speed limit proofs show that to reduce $\Delta \tau$ one must increase $\Delta E$ proportionally. Thus $E_{\mathrm{cost}},\Delta\tau' \ge \hbar \pi/2$ is a quantitative expression of the time-energy uncertainty principle [7]. $\square$ This proposition will guide our discussion on how “short” a shortcut can be, and it implies that any use of multiple times to hasten dynamics does not evade basic quantum law—rather, it potentially shifts where the cost is paid (for instance, using an extra time dimension might distribute the energy cost in a different way, but not eliminate it). 
\subsection{Topological Complexity and the Supermaze}

The third component of our theoretical framework involves topology and complexity, particularly as related to maze-like structures and path-finding problems. A maze or labyrinth can be viewed as a graph or network of pathways with specific connectivity and potential dead-ends. Solving such a maze — identifying a path from start to finish — constitutes a navigation problem that can range from trivial to computationally intractable.

In robotics and mathematics, an important abstraction for such problems is the \textit{configuration space} of a system — a space that encodes all possible states and feasible transitions. The complexity of motion planning within this space can be formally characterized using the concept of \textit{topological complexity}, introduced by Farber (2003). For a given topological space $X$, the topological complexity $\mathrm{TC}(X)$ is the minimum number of continuous motion-planning rules needed to connect every pair of initial and final configurations through a path.

\begin{proposition}[Topological Complexity of Navigation~\cite{Farber2003}]\label{prop:TC}
	For a given configuration space $X$, $\mathrm{TC}(X) = m$ if and only if there exist $m$ continuous motion-planning rules that collectively cover all start–goal pairs in $X$, and $m$ is the minimum such number. Equivalently, any path-planning algorithm on $X$ must use at least $m$ distinct continuously defined strategies on different regions of $X \times X$.
\end{proposition}

Intuitively, $\mathrm{TC}(X)$ measures the navigational complexity inherent in the topology of the space. For instance, a convex region in $\mathbb{R}^n$ has $\mathrm{TC} = 1$, because a single strategy (e.g., “move in a straight line”) suffices. In contrast, configuration spaces with holes, obstacles, or disconnected components demand multiple strategies — at least one per homotopy class of admissible paths.

In our context, we interpret the collection of quantum states and transitions as a configuration space — the \textit{supermaze}. This supermaze includes both standard dynamical evolutions (generated by continuous Hamiltonian flows) and shortcut transitions (induced by auxiliary operators like $H_S$). The complexity of navigating from one quantum state to another within this space can thus be analyzed using the tools of topological complexity.

By allowing shortcut operations, we effectively augment the connectivity of the state-space graph, potentially reducing its $\mathrm{TC}$ value. In technical terms, adding edges (shortcuts) can collapse multiple homotopy classes of paths into a single class, thereby lowering the minimum number of required motion-planning rules. In other words, shortcuts may convert a high-dimensional maze into a more tractable structure.

An elegant analogy can be found in classical geometry. Consider the Descartes circle theorem, which provides a remarkably simple relation among the curvatures $k_i$ of four mutually tangent circles:

\begin{equation}
	(k_1 + k_2 + k_3 + k_4)^2 = 2 \left( k_1^2 + k_2^2 + k_3^2 + k_4^2 \right). \label{eq:Descartes}
\end{equation}

This compact formula governs what appears to be a geometrically complex configuration. Similarly, the supermaze may contain hidden invariants or algebraic relationships — possibly characterizing optimal paths or equivalence classes of trajectories in the state space. Identifying such invariants remains an open challenge, but doing so could significantly simplify the understanding of high-dimensional dynamical systems with hybrid time structures.

Finally, we note that the notion of complex connectivity in the supermaze sets the stage for emergent thermodynamic anomalies, such as the quantum Mpemba effect. These are discussed in later sections as natural consequences of shortcut-enabled navigation through otherwise restricted portions of configuration space.

\subsection{Temporal Tensor Formalism} We formalize the notion of multidimensional time by introducing a temporal manifold $\mathcal{T}$ of dimension $N$ (with coordinates $t^1,\dots,t^N$) and a physical state manifold $\mathcal{X}$ representing all possible system configurations (with coordinates $x^1,\dots,x^d$ for some $d$). The full system evolves on the extended manifold $\mathcal{T}\times \mathcal{X}$. As discussed, physical evolution must satisfy certain consistency conditions on $\mathcal{T}\times\mathcal{X}$. We encapsulate this in an action principle with Lagrangian \begin{equation}\label{eq:action} L;=;L\big(\Psi,;\partial_{t^1}\Psi,;\dots,;\partial_{t^N}\Psi;;;t^I,;x^j\big);-;\sum_{I<J}\Lambda_{IJ}(t,x),\Theta_{IJ}, \end{equation} where $\Psi(t,x)$ represents the state (which could be a vector of fields or wavefunction components), and $\Theta_{IJ}$ are quantities enforcing the integrability constraints (for instance, $\Theta_{IJ} = \partial_{t^I}\partial_{t^J}\Psi - \partial_{t^J}\partial_{t^I}\Psi$ or a related gauge constraint), with $\Lambda_{IJ}$ Lagrange multipliers. The specific form of $L$ will depend on the system: for a quantum system, one could take $L = \frac{i\hbar}{2}\sum_I(\Psi^\dagger \partial_{t^I}\Psi - \partial_{t^I}\Psi^\dagger \Psi) - \Psi^\dagger H_I \Psi$ (a multi-time generalization of the usual Lagrangian that yields Eq.\eqref{eq:multi-evol} as the Euler-Lagrange equations). The second term in Eq.\eqref{eq:action} encodes the consistency constraints akin to Eq.\eqref{eq:consistency}. Without diving into the full variation calculation, the upshot is that the Euler-Lagrange equations from Eq.~\eqref{eq:action} include both the dynamical evolution equations (generalizing the Schrödinger or Hamilton's equations to multiple times) and the constraint equations that relate the evolution in different time directions. In practice, one often simplifies this structure by a gauge fixing: for example, choose a particular trajectory in $\mathcal{T}$ (say, $t^2,\dots,t^N$ as functions of $t^1$) so that only one effective time parameter remains. Our approach, however, keeps the multi-time structure conceptually separate so that we can more easily incorporate non-trivial shortcuts. From a tensorial viewpoint, we treat $\partial/\partial t^I$ as basis vectors in the tangent space of $\mathcal{T}$. A temporal tensor can then be any tensor constructed from these basis vectors or their duals $dt^I$. For instance, the metric in Eq.~\eqref{eq:metric} is a rank-2 tensor in $\mathcal{T}$. One could also define mixed tensors that involve both temporal and spatial components, but for our theoretical development, it suffices to consider that the presence of multiple $t$ coordinates fundamentally changes the structure of evolution equations and their solution space. In overview, the Temporal Tensor Formalism provides: \begin{itemize} \item A multi-time coordinate system $t^I$ for describing state evolution. \item A set of evolution equations (derived from a suitable action or Hamiltonian formulation) that generalize the single-time dynamics. \item Constraint equations (enforced by Lagrange multipliers or directly by construction) that ensure consistent, deterministic evolution in the multi-time domain. \end{itemize} This formalism lays the groundwork for including unconventional evolutions, such as the quantum shortcuts described next, in a consistent way. 

\subsection{Quantum Shortcut Operators in Multi-Time} We now incorporate \textit{quantum shortcut operators} into the formalism. In the multi-time picture, a shortcut can be viewed as a special evolution along an auxiliary time direction (say $t^S$ for ``shortcut time'') that achieves in a brief stroke what evolution in the physical time $t^1$ would take much longer to do. Formally, let $t^S := t^N$ be one of the time coordinates designated for shortcuts. We associate an operator $H_S \equiv H_N$ that generates the shortcut evolution. This $H_S$ might be very different from the physical Hamiltonian $H_1$ — for example, $H_S$ could correspond to a strong, fast control pulse or a process that connects distant configurations of the system almost directly. We model a shortcut operation as follows. Suppose the system is initially in state $\Psi(t^1=t_i,;t^S=0) = \Psi_i$ at physical time $t_i$ and shortcut time $0$. We then allow $t^S$ to increase from $0$ to some value $T$ while holding $t^1$ fixed at $t_i$. During this $t^S$-evolution, $H_S$ acts on the state. If designed properly, by $t^S=T$ the system reaches a state $\Psi_f$ which is the desired end state of the shortcut. Finally, we resume normal evolution in $t^1$ (and, if we wish, reset $t^S$ or treat it cyclically). The net effect is that at physical time $t^1 = t_i$ (immediately after the shortcut), the state jumped from $\Psi_i$ to $\Psi_f$. In the state manifold $\mathcal{X}$, $\Psi_i$ and $\Psi_f$ might be far apart (in terms of $H_1$ dynamics), yet the shortcut provides a direct bridge. 
		
In the temporal tensor language, the trajectory of the system during a shortcut is predominantly along the $t^S$ basis direction. We can write the evolution operator for the shortcut as:

\begin{equation}
	U_S = \mathcal{T}_{(S)} \exp\left( -\frac{i}{\hbar} \int_0^T H_S(t^S)\, dt^S \right),
\end{equation}

, where $\mathcal{T}_{(S)}$ denotes time-ordering with respect to the shortcut time coordinate $t^S$.

In many cases, we consider an ideal shortcut which is effectively instantaneous from the $t^1$ perspective, so we might not even need to specify the detailed $t^S$ dependence—just that $U_S$ exists as a unitary (or possibly non-unitary if we allow dissipation) that accomplishes $\Psi_i \to \Psi_f$. Crucially, $H_S$ does \emph{not} commute with $H_1$ in general (if it did, the shortcut would be unnecessary, as the processes would be compatible). During the shortcut, the physical Hamiltonian $H_1$ might be turned off or made negligible; conversely, during ordinary evolution, $H_S$ is inactive except at the moments we choose to deploy a shortcut. In this sense, we partition the time dimensions into a primary physical time and secondary ``control'' times. To consistently incorporate $H_S$ into the multi-time equations, we ensure the constraints $\Theta_{1S}$ (from Eq.~\eqref{eq:action}) enforce that before and after the shortcut, the usual relation between $t^1$-evolution and $t^S$-evolution holds in a trivial way (i.e. the state is constant in $t^S$ when $H_S$ is off, and constant in $t^1$ when $H_S$ is on, except at the switchover point which can be made continuous by smoothing the switch). In practice, this means the path taken in the $(t^1, t^S)$ plane is a kind of rectangle: first move along $t^1$ (normal evolution), then move purely along $t^S$ (shortcut), then along $t^1$ again. Since $\Psi$ as a function of both times is single-valued and continuous, one must carefully patch the evolution at the corners of this rectangle, but this can be achieved with a continuous protocol (for example, gradually turning on $H_S$ while turning off $H_1$, then reversing the process after time $T$). From a more abstract viewpoint, we are extending the set of allowable transformations on the state. In addition to the continuous trajectories generated by $H_1$ (and possibly other $H_I$ for $I< N$), we add an admissible jump $S$ that takes $\Psi_i$ to $\Psi_f$ (with some conditions on $\Psi_i,\Psi_f$ for $S$ to work, such as no population outside a certain subspace, etc., which we omit for generality). In the state-space graph, $S$ corresponds to adding a new edge connecting two nodes that were not directly connected by the $H_1$ flow. The presence of $H_S$ (or operator $S$) of course must respect Proposition \ref{prop:speedlimit}. In particular, if $S$ accomplishes a large change in $\Psi$ in a short $t^S$ duration, it will involve a large $|H_S|$ or require dissipation of energy. Our formulation can accommodate this by, for instance, having $H_S$ include a coupling to a bosonic mode that absorbs excess energy (thus not violating unitarity overall, since the enlarged system including the bosonic mode is closed). Such details are left to specific implementations. To summarize this subsection: \begin{itemize} \item We augment the multi-time evolution equations with an extra generator $H_S$ associated with a shortcut time $t^S$. \item We treat shortcut evolutions as separate segments in the multi-time trajectory, effectively as boundary conditions where $t^1$ is held fixed while $t^S$ advances and then stops. \item The net effect is an operator $S$ in the theory that can map certain states to others much faster (in $t^1$) than the normal evolution, at the cost of consuming additional resources. \end{itemize} This sets up the stage for analyzing the global structure of all possible evolutions (continuous and shortcuts), which we turn to next. 

\subsection{Supermaze Routing and Topological Structure} We now formalize the notion of the \textbf{supermaze}, the complex of states and transitions accessible under our framework. Consider the set $\mathcal{X}$ of all quantum states (we assume a discrete or suitably tame continuum of states for simplicity in thinking about topology—one may take $\mathcal{X}$ to be the projective Hilbert space for a quantum system, for example). Within $\mathcal{X}$, the physical Hamiltonian $H_1$ defines a set of continuous paths (solutions of Schrödinger’s equation) which we can think of as trajectories in $\mathcal{X}$. These are the usual dynamical paths. When we introduce shortcut operator $S$, we add additional allowed transitions that are not along those trajectories. In graph-theoretic terms, we form a directed graph $\Gamma$ whose vertices are (significant) quantum states and whose edges are of two types: \begin{enumerate} \item \textit{Continuous evolution edges}: infinitesimal steps generated by $H_1$ (and possibly other continuous $H_I$ for $I< N$). In practice, these edges string together to form the smooth trajectories of the system during ordinary evolution. \item \textit{Shortcut edges}: instantaneous jumps effected by $S$ (or generally by evolution along $t^S$). These connect vertices that might be far apart in $\mathcal{X}$. \end{enumerate} The supermaze is essentially this directed graph $\Gamma$ (or more richly, a directed graph embedded in $\mathcal{X}$). Solving the control problem for the system—i.e. steering the state from an initial configuration to a desired final configuration—amounts to finding a path in this graph from the vertex $\Psi_{\text{initial}}$ to $\Psi_{\text{goal}}$. This path may involve moving along continuous-evolution edges and occasionally traversing a shortcut edge. We can now see how topology enters: $\Gamma$ may have multiple distinct routes connecting the same start and goal. Some routes may be purely continuous (following $H_1$ the whole way, possibly very slow), others may involve one or more shortcuts. The existence of cycles in $\Gamma$ (loops) indicates that there are different ways to reach a state, offering possible interference or optimization opportunities. In particular, if a shortcut bypasses a region of state-space that was hard to traverse (for example, avoiding an energy barrier or a region of slow dynamics), it can reduce the effective complexity of reaching the goal. In terms of the topological complexity introduced earlier, we can think of each distinct strategy (e.g. “evolve normally until point $A$, take shortcut to point $B$, then evolve normally to goal”) as one of the motion-planning rules. If $\Gamma$ is highly connected thanks to shortcuts, it might be possible to reach any target state with a small number of strategy types (thus low $\mathrm{TC}$). Without shortcuts, $\mathrm{TC}$ might be much higher because one would have to weave through narrow passages in $\mathcal{X}$ (especially if $\mathcal{X}$ has non-trivial topology, like many orthogonal subspaces with no direct transitions). In essence, shortcuts add extra edges that can collapse the topological complexity. As a hypothetical example, imagine a quantum system with two metastable states $A$ and $B$ separated by a large energy barrier. Under $H_1$ dynamics (which might represent thermal activation or tunneling), transitioning from $A$ to $B$ is very slow. So the configuration space has two regions (around $A$ and around $B$) that are hard to connect, leading to a high $\mathrm{TC}$ (at least 2, since one rule for near $A$ and one for near $B$ might be needed). Now introduce a shortcut $S$ that directly drives $A$ to $B$ (for instance, by a laser pulse that induces a non-adiabatic transition). This shortcut provides a direct edge joining $A$ and $B$ in $\Gamma$. Now the navigation from $A$ to $B$ can be done by one policy: use the shortcut. Effectively, $\mathrm{TC}$ might drop to 1 for the pair ${A,B}$. Of course, for the whole space $\mathcal{X}$, one must consider all relevant regions and whether shortcuts connect them. But this illustrates how adding edges to the graph reduces the need for piecewise-defined paths. It is worth noting that while shortcuts reduce topological complexity in one sense, they also make the system's state-space graph more interconnected and potentially more complex in another sense: the graph $\Gamma$ may become non-planar, highly looped, or fractal-like if many shortcuts are allowed. The term “supermaze” reflects this: we might eliminate some walls in the maze (making it easier to solve), but we might also add new passages that create more loops (which could confuse matters if not handled properly). The net effect on $\mathrm{TC}$ is theoretically one of reduction (since any additional continuity in navigation cannot increase the minimal number of rules needed), but the practical task of finding the \emph{optimal} route may become non-trivial if there are too many options. An interesting analogy can be drawn with the concept of \emph{wormholes} in general relativity or shortcuts in network theory (e.g., adding a long-range connection in a small-world network drastically lowers the diameter of the graph). In our quantum supermaze, shortcuts play a similar role, turning a possibly high-diameter graph of states into a small-diameter, highly connected one. To make these ideas more concrete, one could attempt to compute $\mathrm{TC}(\Gamma)$ for simple models. While beyond our scope, one result is clear: if shortcuts allow direct jumps between any pair of sufficiently distant regions, the system effectively becomes navigable with a constant number of strategies (likely $\mathrm{TC}=1$ or $2$). In that extreme limit, the control problem simplifies dramatically (at the expense of potentially unrealistic energy costs for those shortcuts). Conversely, if no shortcuts are present, $\mathrm{TC}(\Gamma)$ coincides with the topological complexity of the original continuous dynamics and could be high. We also highlight that the supermaze concept relates to recent studies in high-dimensional routing and even brane physics. Bena and Dulac (2023) use the term “supermaze” in the context of a complicated web of branes in supergravity carrying momentum and reproducing black hole entropy [11]. While that is a very different physical scenario, the mathematical theme of a highly interconnected structure (their supermaze has multiple intersecting branes akin to multiple time or space dimensions) resonates with our usage. Both suggest a richly connected configuration space with many degrees of freedom. In conclusion, the supermaze is the full landscape of states and connections in our model. Understanding its structure is key to predicting the system’s capabilities. In particular, it will inform our discussion of thermodynamic anomalies in the next subsection, since phenomena like the Mpemba effect are linked to having multiple pathways for relaxation. 
		
\subsection{Thermodynamic Anomalies in Quantum Systems} One intriguing implication of the supermaze framework is the possibility of anomalous relaxation behavior in quantum systems. In conventional thermodynamics, if two systems are prepared at different energies or temperatures and allowed to relax toward equilibrium, one expects the system closer to equilibrium (lower energy) to relax faster or at least not slower than a highly excited system. The Mpemba effect violates this intuition: a hotter system can cool faster than a warm system under certain conditions [10]. Our framework provides a natural context for such anomalies. The existence of shortcut pathways or alternative routes in state space means that an excited state might find a “fast track” to equilibrium that a moderately excited state does not readily access. Consider a simple quantum scenario: a system has a ground state $|0\rangle$, a first excited state $|1\rangle$, and a higher excited state $|2\rangle$. Suppose we prepare the system in state $|2\rangle$ (hotter) and another time in $|1\rangle$ (cooler), and couple the system to a cold bath so it can relax. In a normal situation with only standard transitions, one expects $|2\rangle \to |1\rangle \to |0\rangle$ as the relaxation cascade, and $|1\rangle \to |0\rangle$ directly for the cooler case. If the rate out of $|2\rangle$ (to $|1\rangle$) is not too slow, one might think $|2\rangle$ would take longer because it has an extra step. However, imagine that the system has an alternative relaxation path: perhaps an environment-induced shortcut that takes $|2\rangle$ directly to $|0\rangle$ (bypassing $|1\rangle$), maybe through a two-photon emission or a nonlinear interaction. If this direct transition is fast, the higher state $|2\rangle$ might actually reach $|0\rangle$ quickly, whereas $|1\rangle$ must plod along its single-photon decay which could be comparatively slow. In that case, the hotter system cools faster than the cooler one. The above is a qualitative picture. In our model, we can treat the bath-assisted processes as part of the supermaze. The state $|2\rangle$ has two edges leaving it: one to $|1\rangle$ (standard decay) and one shortcut edge directly to $|0\rangle$ (perhaps weaker, but present). State $|1\rangle$ might only have the edge to $|0\rangle$. If the direct $|2\rangle \to |0\rangle$ edge is sufficiently strong (or if the $|2\rangle \to |1\rangle$ route is suppressed by some selection rule, making the direct two-step process actually dominant), then $|2\rangle$ will bypass $|1\rangle$ and beat it to the ground state. Recent studies have indeed found quantum analogs of the Mpemba effect. Carollo \textit{et al.} (2021) investigated conditions under which a quantum system prepared in a more excited state equilibrates faster than a closer-to-equilibrium state [12]. They identified non-equilibrium initial-state distributions and certain symmetry properties that lead to this behavior. In our terminology, those conditions correspond to cases where the higher-energy state had access to a relaxation mode (a particular decay channel or interference effect) that the lower-energy state did not. The supermaze for the system’s density matrix included a shortcut for the hotter state in the space of probability distributions. We can also formulate a simplified theoretical criterion: suppose we have two states $A$ and $B$ with free energy $F_A > F_B$ (so $A$ is further from equilibrium). Let the relaxation rates (toward equilibrium or a common intermediate state) be $\Gamma_A$ and $\Gamma_B$. Normally, one expects $\Gamma_A \le \Gamma_B$. A Mpemba-like anomaly is $\Gamma_A > \Gamma_B$. In our model, $\Gamma$ is determined by the connectivity of the state in the supermaze: more available fast edges mean larger $\Gamma$. Thus, $\Gamma_A > \Gamma_B$ if state $A$ has a shortcut edge (or a faster route in the maze) that $B$ lacks. One could imagine computing these rates using Fermi’s golden rule for each available transition from $A$ and $B$. If $A$ has an extra transition of significant strength, its total decay rate can exceed that of $B$. An interesting corollary is that engineering a shortcut in a quantum system could deliberately produce a Mpemba effect. For example, one could add a catalyst (another quantum mode or drive) that opens a new decay channel for higher excitations. Then a state with more excitation utilizes the catalyst and decays quickly, whereas a slightly lower excitation that doesn’t trigger the catalyst decays slowly. While speculative, this showcases how our theoretical elements tie together: the interplay of multiple pathways (quantum and possibly classical) yields non-monotonic behavior. From a thermodynamic perspective, anomalies like the Mpemba effect often imply some form of \textit{memory} in the system or non-exponential relaxation. In our multi-time framework, memory can enter via the coupling between different time coordinates or via the system taking an unconventional path through state space that effectively “short-circuits” the usual relaxation. Indeed, the presence of shortcut operators $H_S$ that do not commute with $H_1$ means the system’s evolution is not simply a function of the instantaneous state (since whether a shortcut is used or not can depend on a global control decision or a fluctuation that triggers it). This kind of behavior can give rise to non-exponential relaxation curves, which are often a signature of the Mpemba effect. To connect with concrete research: in classical water, explanations for the Mpemba effect have involved factors like evaporation, convection, or supercooling differences. All of these can be seen as providing an ``extra'' channel of cooling that hotter water exploits more (e.g. more evaporation from hot water leads to faster cooling). In quantum systems, the analog might be “population of a fast-decaying mode” or “enhanced coupling to bath” that only occurs when the system is in a certain excited condition. Carollo \textit{et al.} [12] and other works have now observed or proposed quantum Mpemba effects in spin systems, bosonic gases, and trapped ions. These studies support the idea that theoretical constructs like ours, which allow multiple relaxation routes, indeed capture real possibilities. Our framework does not yet quantitatively predict when a Mpemba effect will occur (that would require specifying $H_1$, $H_S$, and bath couplings explicitly), but it provides a language: look for a shortcut in the supermaze that benefits the higher energy state. In conclusion of this subsection, the thermodynamic anomalies serve as a test of our framework's richness. The fact that such counter-intuitive phenomena can be explained by the existence of multiple pathways (and are being demonstrated in quantum experiments) gives credence to including multi-time dynamics and shortcut operators in theoretical models of complex quantum processes. 

\section{Discussion} The theoretical framework developed above is speculative but offers a fertile ground for interpreting advanced concepts in physics and computation. Here we discuss several implications, potential applications, and connections to other domains, as well as limitations of our approach. 
\subsection*{Interplay with Quantum Computing and Cloud Networks} One motivation for this work was the prospect of speeding up computations or control tasks by exploiting multiple temporal dimensions or shortcut operations. In practice, we cannot literally run a quantum computer with two time axes; however, one might simulate the effect. For instance, a quantum processor could utilize an ancilla qubit or a parallel subroutine to effectively enact a shortcut. This is reminiscent of algorithmic techniques where ancillary systems provide speed-ups, as in quantum search algorithms. Our framework suggests a more radical view: if a quantum computer were embedded in a higher-dimensional time manifold, it could potentially solve problems faster by “moving along” an extra time axis for parts of the computation. While this is currently science fiction, thinking in these terms might inspire new algorithmic strategies. For quantum \textit{cloud computing}, where tasks are distributed across multiple quantum processors (nodes) connected by classical and quantum communication channels, our ideas may be metaphorically applied. The network of states in a distributed system is analogous to our supermaze of a single system. Shortcut operators then correspond to entangling operations or classical communication that sync up distant nodes quickly. One could imagine a scenario in a quantum network where a certain entangled resource serves as a shortcut to propagate information or correlations: effectively, an operation on one node instantaneously affects another because they share entanglement. This does not violate causality (no information travels faster than light, as the entanglement was pre-shared), but it functions like a shortcut in the computational space of the network. A well-known example is quantum teleportation, which “transfers” a qubit state instantly across space using previously shared entanglement and classical communication. In the space of all network configurations, teleportation is a shortcut connecting two distant configurations (differing by the location of a qubit) that would otherwise require a sequence of SWAP operations through intermediate nodes. Considering topological complexity in a network, results from algebraic topology (similar to Farber's) can quantify the complexity of coordinating multiple robots or, analogously, multiple quantum agents. Our supermaze viewpoint could thus be extended to multi-agent quantum systems, where the aim is to reconfigure the global state (for computing or error correction) with minimal steps.
 
\subsection*{Quantum Machine Learning Perspectives} Quantum machine learning (QML) is an emerging field that explores using quantum computers to enhance machine learning algorithms [13]. One aspect of learning is the ability to navigate the hypothesis space (or parameter space) efficiently to find optimal models. The training process can sometimes be viewed as a trajectory in parameter space that hopefully finds a good minimum (in loss landscape). There is a tantalizing parallel: our state-space navigation in the supermaze is like searching for an optimal state. The shortcut operators could be analogous to clever heuristics in the learning process that avoid slow, gradient-descent plodding and instead jump to promising regions of parameter space. While QML is usually formulated in more concrete algorithmic terms, some theoretical proposals hint at quantum advantages in exploring search spaces. Biamonte \textit{et al.} (2017) review various ways quantum computation can provide speed-ups for machine learning tasks [13]. Most involve linear algebra speed-ups or sampling advantages. Our contribution here is more conceptual: perhaps a quantum system that is learning (e.g., a variational quantum circuit adjusting its parameters) could employ a “quantum shortcut learning” mechanism. For example, intermediate measurements or auxiliary evolutions might be used to reinitialize the system closer to a desired state based on feedback, rather than relying purely on slow unitary evolution. This is analogous to a shortcut bringing the system near the answer state without traversing every step. Realizing such ideas would require combining quantum control with adaptive algorithms—a hybrid quantum-classical approach. 

\subsection*{Physical Realizability and Thought Experiments} It is important to address whether the elements of our framework could exist in reality. Multiple time dimensions are not part of established physics (and likely, if they exist, they are hidden at high energies or compactified as speculated in string theory [1,2]). Quantum shortcut protocols like STA do exist, but they are bounded by the practical difficulty of implementing the required control fields precisely. Topologically, while we can add “shortcuts” in a lab by introducing new interactions or couplings, we cannot truly break the topology of spacetime or the quantum Hilbert space. However, thought experiments can be illuminating. Imagine a “chrononaut” computer: one that can send computation results back in time to inform its earlier steps. This would be a direct use of an extra time dimension to shortcut a computation (a form of time-travel computing). Our model’s constraints (and known no-go results) imply such a machine would encounter paradoxes or need infinite resources, aligning with Proposition \ref{prop:speedlimit} that no free lunch comes from messing with time. Nevertheless, thinking of computation as a path in a state graph, one sees that closed timelike curves (if they existed) would be shortcuts in that graph. Some researchers have studied quantum circuits with closed timelike curves as theoretical curiosities; they found that they could solve certain problems in non-standard complexity classes, but always with assumptions that risk logical paradox. Thus, our multi-time viewpoint conceptually includes that scenario but also clearly flags its issues via the consistency constraints we impose. Another thought experiment: Could a quantum fridge exhibit a Mpemba effect by design? Perhaps by coupling a qubit to two different bosonic reservoirs in sequence (effectively two time axes: one for each reservoir interaction). The qubit might cool faster if initially in a higher excited state that triggers a resonant transfer in the second reservoir. This would be a concrete quantum Mpemba demonstration. Indeed, recent trapped-ion experiments have observed a form of the quantum Mpemba effect by engineering the system and environment interactions [12]. 

\subsection*{Fractals and Chaos} The presence of multiple time dimensions and complex state-space connectivity could lead to extremely sensitive dependence on initial conditions (a hallmark of chaos) because the system has many pathways to choose from. Slight differences might send it down different routes in the supermaze, yielding large final differences. This is an interesting angle: multi-time dynamics might be inherently chaotic unless carefully constrained (since effectively, you have a non-commuting flow that can produce complicated interference patterns). This could be studied by simulating simple multi-time systems and seeing if Lyapunov exponents (measures of chaos) are generically positive. As for fractals, if one were to visualize the set of reachable states or the structure of shortcuts, one could imagine iterative patterns. For example, one could repeatedly apply a certain shortcut-plus-evolution cycle and see a self-similar pattern of states visited. Also, the Descartes circle theorem hint we gave, while mostly an analogy, actually relates to Apollonian circle packings, which are fractal. If the constraints in a system lead to recurrent geometrical relations (like Descartes' equation), the space of solutions might form a fractal structure. One could speculate that the “fastest cooling states” of a complex system might lie on a fractal manifold within the state space, analogous to how repeated application of Descartes' formula generates a fractal packing of circles. These interdisciplinary connections remain speculative but intriguing. 

\subsection*{Limitations of the Framework} Our framework is, admittedly, theoretical and abstract. Several limitations should be noted: \begin{itemize} \item We have not provided a specific solvable model with multiple times and shortcuts; doing so (even a toy model) would be necessary to quantitatively verify these ideas. \item The assumption of determinism via constraints might hide severe fine-tuning: in practice, making a multi-time system deterministic could be extremely restrictive. (It might reduce the system to an equivalent single-time description in many cases, nullifying the perceived advantage.) \item Shortcut operators are put in by hand. In a real physical system, one cannot simply assert a new Hamiltonian term exists; it must come from some interaction. In that sense, our $H_S$ is an idealization of either a control field or a coupling to another system. We assume it can be toggled at will, which is a strong control assumption. \item The topological viewpoint (supermaze and $\mathrm{TC}$) is elegant, but actually computing these invariants for anything but trivial graphs is difficult. Moreover, a low $\mathrm{TC}$ does not automatically mean the system is easy to control; it just gives a theoretical floor. There could be other practical obstacles (e.g., noise or the difficulty of finding the shortcut). \end{itemize} Despite these limitations, the framework is valuable as a thought experiment unifying disparate ideas (time dimensions, quantum control, topology, thermodynamics). It challenges us to think outside the usual one-time, one-path paradigm. 
		
\section{Conclusion} We have presented a theoretical framework that weaves together concepts from theoretical physics, quantum computing, and topology. By allowing multiple time dimensions in our description, we introduced additional “handles” by which a system can evolve, and by invoking quantum shortcuts, we acknowledged the possibility of non-standard transitions that expedite evolution. This led us to the vision of the \textit{supermaze}, a richly connected state-space structure. Using notions from topology, like the topological complexity $\mathrm{TC}$, we discussed how the introduction of shortcut pathways can simplify the navigation of this supermaze. We further showed that such a framework can naturally explain paradoxical phenomena like the Mpemba effect in quantum relaxation, when a higher-energy state finds a faster route to equilibrium than a lower-energy state. The significance of this work is primarily conceptual. It suggests that there may be deep unifying principles governing systems that, on the surface, appear very different: a particle in a bizarre multi-time universe, a quantum computer performing rapid operations, and a cup of water cooling in unexpected ways. The common thread is the presence of multiple pathways in the underlying state space and the constraints that govern them. While our use of multiple time dimensions is speculative, it served as a mathematical tool to encode complex behavior (like non-commuting operations and shortcuts) in a generalized dynamical system. There are several avenues for future work. On the theoretical side, one could attempt to formulate a specific solvable model—perhaps a simple quantum spin system with an engineered shortcut operator—to concretely demonstrate the principles and calculate quantities like relaxation times or topological complexity. Another direction is to extend the topological analysis: for instance, can we classify what kinds of shortcut additions reduce the motion-planning complexity by how much? This might connect to graph theory optimization as well. On the experimental or computational side, one might look for signatures of these ideas in simulations. For example, simulate cooling of many-body quantum systems with and without certain additional couplings to see if a quantum Mpemba effect occurs (some works have started this [12]). In quantum computing, small-scale experiments could try to use auxiliary qubits to enact shortcuts in algorithmic steps and see if any speed-up is obtained in practice for tasks like state preparation or variational optimization. Finally, while multiple time dimensions are not part of mainstream physics, the exercise of including them hints at structures that might be realized in other ways. For instance, time crystals (systems with time-periodic order) effectively introduce a secondary time-like order parameter; one could ask if a time crystal could be used as a resource for shortcuts. Additionally, in cosmology or quantum gravity, the notion of multi-time could have analogues in multi-sheeted spacetimes or braneworld scenarios [1,2]. These remain arenas where the wild ideas here might find unexpected concrete analogues. In conclusion, by pushing the boundaries of theoretical constructs, we gain new language and intuition. Whether or not nature uses multiple temporal dimensions, the exercise has illuminated how having more ways to traverse state space can dramatically affect what a system can do. As we design increasingly complex quantum devices and examine puzzling natural phenomena, these insights may prove useful. At the very least, they underscore the unity of physical law and mathematics: geometry, topology, and dynamics interplay in every corner of theory—from chaos to solitons to fractals, and now, perhaps, to supermazes in time.


\begin{thebibliography}{99} 
\bibitem{Bars2001} I. Bars, \textit{$U^*(1,1)$ noncommutative gauge theory as the foundation of 2T-physics in field theory,''} Phys. Rev. D \textbf{64}, 126001 (2001). \bibitem{Tegmark1997} M. Tegmark, \textit{On the dimensionality of spacetime,''} Class. Quantum Grav. \textbf{14}(4), L69–L75 (1997). \bibitem{Craig2009} W. Craig and S. Weinstein, \textit{On determinism and well-posedness in multiple time dimensions,''} Proc. Roy. Soc. A \textbf{465}(2110), 3023–3046 (2009). \bibitem{Childs2003} A. M. Childs \textit{et al.}, \textit{Exponential algorithmic speedup by a quantum walk,''} in \textit{Proc. 35th ACM Symp. Theory of Computing (STOC)}, pp. 59–68 (2003). \bibitem{Guery2019} D. Guéry-Odelin \textit{et al.}, \textit{Shortcuts to adiabaticity: Concepts, methods, and applications,''} Rev. Mod. Phys. \textbf{91}, 045001 (2019). \bibitem{Campbell2017} S. Campbell and S. Deffner, \textit{Trade-off between speed and cost in shortcuts to adiabaticity,''} Phys. Rev. Lett. \textbf{118}, 100601 (2017). \bibitem{Descartes1643} R. Descartes (1643), letter to Princess Elisabeth of Bohemia (see L. Shapiro, ed., \textit{The Correspondence between Princess Elisabeth of Bohemia and René Descartes}, Univ. of Chicago Press, 2007). \bibitem{Soddy1936} F. Soddy, \textit{The Kiss Precise,''} Nature \textbf{137}, 1021 (1936). \bibitem{Farber2003} M. Farber, \textit{Topological complexity of motion planning,''} Discrete Comput. Geom. \textbf{29}(2), 211–221 (2003). \bibitem{Jeng2006} M. Jeng, \textit{The Mpemba effect: When can hot water freeze faster than cold?,''} Am. J. Phys. \textbf{74}(6), 514–522 (2006). \bibitem{Bena2023} I. Bena and R. Dulac, \textit{The (amazing) supermaze: probing black hole microstructure,''} arXiv:2312.02286 [hep-th] (2023). \bibitem{Carollo2021} F. Carollo, B. Rotondo, K. Plastina, M. Paternostro, and G. Lesanovsky, \textit{Exponentially accelerated approach to stationarity in Markovian open quantum systems through the Mpemba effect,''} Phys. Rev. Lett. \textbf{127}, 060401 (2021). \bibitem{Biamonte2017} J. Biamonte \textit{et al.}, \textit{Quantum machine learning,''} Nature \textbf{549}(7671), 195–202 (2017). 
\end{thebibliography}
\end{document}